\documentclass{elsart}
\usepackage{amssymb}
\usepackage{graphicx}
\begin{document}
\begin{frontmatter}
\title{\Large \bf Complexation in  Polyelectrolyte Solution with  Divalent  Surfactant  }
  \author[dfte]{Marcelo B. da Silva}
  \ead{bruno@dfte.ufrn.br}
  \author[dfte]{Liacir S. Lucena}
   \ead{liacir@dfte.ufrn.br}
  \author[ufrgs]{Marcia C. Barbosa}
  \ead{barbosa@if.ufrgs.br}
  \ead[url]{ http://www.if.ufrgs.br/$\sim$barbosa}
  \address[dfte]{Departamento de F\'{\i}sica Te\'orica e
Experimental and International Center for Complex Systems, Universidade Federal do Rio Grande do Norte,
Natal, RN,  CEP 59078-970, Brazil }
  \address[ufrgs]{Instituto de F\'{\i}sica, Universidade Federal do Rio
  Grande do Sul\\ Caixa Postal 15051, 91501-970, Porto Alegre,
  RS, Brazil}
  \begin{abstract}

     We study a simple model of DNA divalent cationic surfactant complexation.
We find that the combination of electrostatic and
     hydrophobic effects leads to a cooperative phenomenon  in which
  as the amphiphile is added to
                  the solution containing DNA,  a large fraction of the DNA's charge is
     neutralized by the condensed divalent  cationic surfactants, forming the surfoplex.
This binding transition occurs for concentrations that are  lower for divalent than for monovalent surfactants. Since the electrostatic strength is larger in the first case and the amount of surfactant lower, we suggest that
multivalent amphiphilic molecules would be more efficient than monovalent  for transfection.

\end{abstract}

  \end{frontmatter}

\maketitle

\section{Introduction}

      Gene therapy represents a promising way for
      the treatment of both genetic and acquired diseases.
      The basic idea is to replace the sick gene with a healthy
      one or in some cases to add a new gene to get a resulting synthesis
      of a therapeutic protein.
      The process of getting the new gene to its target
      involves the crossing of several barriers, such as the cell
      membrane and the nuclear membrane.
      Since both the DNA and the cell membranes are negatively charged,
the naked polynucleotides are electrostatically prevented from entering
into the cell.
  Viral vectors such
  as retroviruses and adenoviruses are very efficient
  and able to target a wide range of cells \cite{Gr85}\cite{Be93}.
  The gene transfection is accomplished  by the use of
  a virus in which the native DNA has
  been replaced by the required DNA.
  Since the main purpose of the virus is to replicate
  itself, the new gene
  is successfully transfected to the cell nucleus by endocytosis or
  membrane fusion. This process suffers from the drawback that
  in cases where repeated treatment is needed, this might cause the
  immune system to react negatively due to the viral
  origin of the vector.

  As a solution for this problem nonviral vectors have been developed.
  One of the approaches, pioneered by Felgner and Ringold
\cite{Fe89}\cite{Fe91}
relies on the association between anionic nucleic acid and cationic
lipid
lipossomes. The process of association neutralizes the excess negative
charge
of the DNA and the DNA-lipossome penetrate into the cell by endocytosis.
The efficiency
of this method is low and they are toxic to the cell at
the concentrated used.

Among the parameters needed for achieving efficient transfection
is the requirement that the complex should have small size similar
to a virus \cite{Pi97}. This factor indeed limits the efficiency of
the DNA-lipossomes complexes. In contrast, cationic surfactant
have been shown to condense to DNA into discrete particles
containing a single nucleic acid molecule \cite{Go98}\cite{Sh87}
that can neutralize and revert the charge of the DNA
\cite{Ku98}-\cite{Si01} allowing the surfoplex to approach the
membrane. Monovalent DNA-surfactant complexes exhibit a discrete
first-order phase transition between  elongated coil and collapsed
globule \cite{Me95} for a concentration of amphiphilic molecules
well below the miscelar concentration. Despite this unique
feature, monovalent detergent are poorly efficient in vitro in
gene transfer \cite{Pi89}\cite{Ro91}. When the surfoplex
approaches the membrane, the interaction between the surfactant
molecules and the phospholipid bilayer overcomes the electrostatic
attraction between the anionic nuclei acid and the cationic
detergent. The surfactant incorporate into the phopholipid
membrane resulting in the unfolding of the DNA that stays outside
the cell \cite{Cl00}.

Therefore, in order to have a stable surfoplex, one needs to
enhance the electrostatic interaction. In this paper we
present a model of DNA-amphiphilic solution where the
surfactant is divalent. We find that in equilibrium, solution
consists of complexes composed of DNA and
associated counterions and amphiphiles. Even for an small
amount of amphiphiles in the solution, the cooperative binding is
found. Due to the high valence, the interaction between
the nuclei acid and the detergent is expect to overcome 
 the interaction between the surfactant molecules and the
phospholipids
in the membrane and stay associated to the DNA allowing it 
to transpass the membrane. The complex formed is more stable than the monovalent
structure, forming a  collapsed globule that we expect
will transect with high efficiency \cite{Ka02}.

 \section{The Model}

   Our model, illustrated in Fig. 1, consists of a solution of
    {\it DNA} strands  of length $L$ and   diameter
    $a_{p}$, divalent surfactant and monovalent salt. In aqueous
solution, the polyions become ionized
resulting in a negative charge   $-Zq$ distributed uniformly distributed
with separation $b=L/Z$.   The solvent, water,  is
  modeled  as a continuous medium of
  dielectric constant $D$. The ions of the salt are
  completely dissociated, forming an equal number
  of positive and negative ions. Similarly,
   the surfactants are assumed to be fully dissociated producing
   negative monovalent coions and polymeric chains with
   a divalent cationic head group. For simplicity, all the counterions
   and coions are  treated as identical, independent of the molecules
   from which they were derived. The electrolytes are  depicted as
   hard spheres of diameter $a_c$ and charge $\pm q$ and
   the surfactant is modeled as a polymer of $s$ monomers each one
   considered as a rigid sphere of diameter $a_c$ with the head monomer carrying
   a charge of $+2q$.

\begin{figure}

\includegraphics[width=8cm,height=6cm]{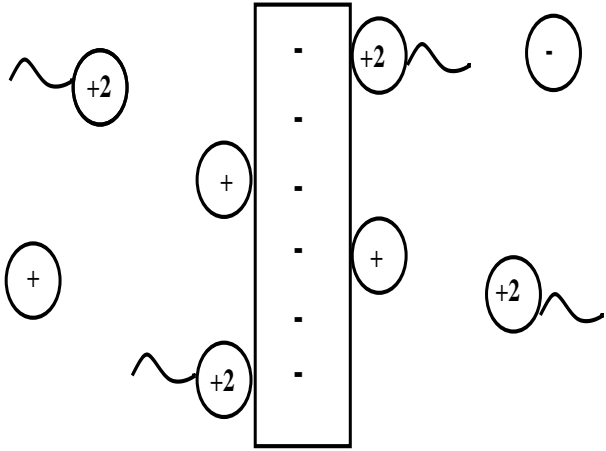}
\caption{The model}
\label{fig:fig1}

\end{figure}
   The interaction between the hydrophobic tails is short ranged and
   characterized by the parameter $\chi$.
   The density of DNA strands is
   $\rho_{p}=N_p/ V$,  the density of
   monovalent salt is  $\rho_{s}=N_s/V$ and the density of
   divalent amphiphiles is $\rho_{a}=N_a/V$.

   The strong electrostatic interactions between the polyions, the
   counterions, salt cations and surfactant heads leads to the
   formation of complexes, which in thermodynamic equilibrium
   will be made up of one polyion, $n_c$ monovalent counterions and
   $n_a$ divalent surfactant. We do not consider the effects of
   of polydispersity in the size of the complexes, since it does
   not affect the final result. Due to the association and to the
   charge conservation, there are only two free quantities and so,
   \begin{eqnarray}
   \rho_{c}=(Z-n_{c})\rho_{p}+\rho_{s} \; \; \;\;
   \rho_{a+} =\rho_{a}-n_{a}\rho_{p} \;\; \;\;
   \rho_{-}=\rho_{s}+2\rho_{a} \label{eq1}
   \end{eqnarray}
   where $\rho_c$ is the density of free monovalent counterions,
$\rho_{a+}$
is the density of free amphiphiles and $\rho_-$ is the density of
negative ion.

The objective of this theory is to determine the number of
counterions $n_{c}$ and surfactants $n_{a}$ associated to each {\it DNA}
strand. For this, we construct the Helmholtz free enery  of the
system and minimize it. The details about the model
can be found elsewhere \cite{Ku98}-\cite{Si01}. We give
here the main steps.
The relevant contributions for the Helmoltz  free
energy  are two,  the electrostatic and the entropic namely:
\begin{equation}
F= F_{el} + F_{ent} \label{eq2}
\end{equation}

In $F_{el}$ three types of interactions
can be found: between the free ions and free
surfactants ,$F_{is}$, between  the complex, free ions
and free surfactants ,$F_{pis}$, and between the  complexes  $F_{pp}$.
\begin{equation}
F_{el}= F_{is} + F_{pis} + F_{pp} \label{eq3}
\end{equation}

With the aid of the theory of Debye-H\"uckel-Bjerrum (DHBj) it is
possible to find the electrostatic interaction
between  the complexes, ions and
surfactant given by \cite{Ku98}-\cite{Si01}:

\begin{eqnarray}
\beta f^{pis}=-\frac{\rho_{p}Z_{c}^{2}(a/L)}{T^{\ast }(\kappa a)^{2}}
\left\{ 2\ln \left[ \kappa aK_{1}(\kappa a)\right] -I_{0}+
                                   \frac{(\kappa
a)^{2}}{2}\right\} \,  \label{eq4}
\end{eqnarray}
with $\beta=1/k_BT$ and
\begin{equation}
I_{0}\equiv \int_{0}^{\kappa a}\frac{xK_{0}^{2}(x)}{K_{1}^{2}(x)}dx \label{eq5}
\end{equation}
where $\kappa$ in $(\kappa a)^2=4\pi \rho_1^*/T^*$ is the inverse of the Debye
screening length, $\rho_1^*=\rho_1
a^3=\rho_{c}a^3+4\rho_{a+}a^3+\rho_{-}a^3$ is the reduced density and
$T^*=Dk_BTa/q^2$ is the reduced temperature. Furthermore,
$Z_c=Z-n_c-2n_a$ is the valence of each complex and
$a=(a_c+a_p)/2$ is the effective radius of the exclusion cylinder
around each complex.

In the framework of the Debye-H\"uckel theory, the interaction between
the free ions is given by:
\begin{eqnarray}
\beta f^{is}=-\frac{1}{4\pi a_{c}^{3}}\left[\ln
(1+\kappa a_{c})-\kappa a_{c}+\frac{(\kappa
a_{c})^{2}}{2}\right] \label{eq6}\; .
\end{eqnarray}

The electrostatic free energy interaction between two complexes for
large separations is
screened. The short-range of this interaction allows  one to use a
mean-field approximation resulting in:
\begin{equation}
\beta f^{pp}=\frac{2\pi a^{3}Z_{c}^{2}\rho_{p}^{2}\exp (-2\kappa a)}{%
T^{\ast }(\kappa a)^{4}K_{1}^{2}(\kappa a)}\,.  \label{eq7} \;
\end{equation}

The calculation of the entropic contribution can be obtained with
the aid of Flory
\cite{Fl71} theory of mixing. The free energy is
a sum of ideal free energies of
various species, namely
\begin{equation}
\beta f^{ent}=\sum \left[ \rho_{s}-\rho _{s}\ln \left( \frac{\phi_{s}}{%
\zeta_{s}}\right) \right]   \label{eq8}
\end{equation}
where $s$ represents the different species and $\zeta_{s}$ is the
internal partition of the species $s$. In the case of the
particles without structure, the internal partition function
$\zeta_ {-} = \zeta_{c} = \zeta_{a+} = 1$. The volume fraction
$\phi_{s}$  of the different species are:
\begin{eqnarray}
\phi_{p}&=& \frac{\pi \rho_{p}^{\ast }}{4(a/L)}\left( \frac{a_{p}}{a}%
\right)^{2}+\frac{Z\pi \rho_{p}^{\ast}}{6}(s_{a}m_{a}+m_{c})\left( \frac{a_{c}}{a}\right)^{3}\nonumber \\
\phi_{c}&=&\rho_{c}^{\ast }\frac{\pi }{6}\left(\frac{a_{c}}{a}\right)^{3} \nonumber \\
\phi_{a+}&=&\frac{s_{a}\pi \rho_{a+}^{\ast }}{6}\left( \frac{a_{c}}{a}%
\right)^{3}\,  \nonumber  \\
\phi_{-}&=&\frac{\pi \rho_{-}^{\ast
}}{6}\left( \frac{a_{c}}{a}\right)^{3}\; . \label{eq9}
\end{eqnarray}
Here we introduce the fractions counterions, $m_c=n_a/Z$, and
surfactant, $m_a=n_c/Z$, associated to each DNA strand.
Accounting for all these terms, the entropic contribution becomes
\begin{eqnarray}
\beta f^{ent} &=& \rho_{p} \ln \left(
\frac{\phi_{p}(1+m_{c}+m_{a})}{\zeta_{cl}(1+m_{c}+m_{a}s_{a})}\right) -
\rho_{p}  \nonumber \\ &+&\rho_{c}\ln\phi_{c}-\rho_{c}+
\rho_{a_{+}}\ln\frac{\phi_{a_{+}}}{s_{a}}-\rho_{a_{+}} \nonumber
\\ &+&\rho_{-}\ln\phi_{-}-\rho_{-} \label{eq10}
\end{eqnarray}
where the entropy for the free  surfactants arises from the Flory
theory for polymers
\cite{Fl71} which  is also the basis of the  entropic contribution
for the complex \cite{Ku98}. The internal partition
function of the complex, $\zeta_{cl}$,  can be calculated by modeling the DNA
by an one dimensional lattice with Z sites. If the number of
associated ions to each site can be only zero or one,
this problem becomes equivalent to finding the free
energy  of an
one dimensional array with the
three different states.

Due to the presence of two different valences, the exact solution
of this model is not trivial, so we employ the mean-field
Gibbs-Bogoliubov-Feymman inequality. The resulting   partition
function is given by
\begin{eqnarray}
-ln\zeta_{c}\left[ m_{c},m_{a}\right] &=&\xi K\left[ \frac{%
Z_{c}}{Z^{2}}^{2}-1\right]
+\beta \chi
\left(Z-1\right) m_{a}^{2} \nonumber \\
&+&Zm_{c}\ln m_{c} + Zm_{a}\ln m_{a} \\  \nonumber 
&+&Z\left(1-m_{c}-m_{a}\right)\ln \left( 1-m_{c}-m_{a}\right)
\label{eq11}
\end{eqnarray}
where $\xi \equiv \beta q^{2}/Da$ is the Manning parameter,
$K=Z[\psi(Z)-\psi (1)] -Z+1 $, and $\psi(n)$ is the digamma function.

The equilibrium configuration of the system is found by the
minimization of the Helmoltz free energy, leading to two
equations, namely
\begin{eqnarray}
\frac{\partial F}{\partial m_{c}}\delta m_{c} &=&0 \nonumber \\
\frac{\partial F}{\partial m_{a}}\delta m_{a} &=& 0  \; .
\label{12}
\end{eqnarray}
Solving this system of two equations, it is possible to obtain the
                              values of $m_{c}$ and $m_{a}$.

\section{Results and Conclusions}


We define a "surfoplex" to be a
     complex in which almost all of the DNA's phosphate groups are neutralized by the associated surfactant molecules.As mentioned earlier, we are interested in the minimum amount of cationic surfactant needed to transform naked DNA
     into surfoplexes. To this effect, we study the dependence of the number of condensed surfactant molecules  on
     the bulk concentration of surfactant $\rho_a$. In order to evaluate the relevance of hydrophobic interactions between the amphiphiles, the
    hydrophobic parameter  was varied from 0 to $\beta\chi= -6$. The effect
of addition of high concentrations of salt to the system was
analyzed by varying the
amount of salt added to the system $\rho_s$.

Fig.$2$  and Fig.$3$ illustrate the surfactant and the counterion
binding isotherm, $m_a$ and $m_c$ respectively,  as a
function of total amphiphilic
     concentration for two different hydrophobicity parameter
$\beta\chi=-3.5,-6$ and salt concentrations $\rho_s=10^{-3}M,10^{-4}M$.
At very low densities of divalent surfactant, the condensation is
     dominated by the monovalent ions, since the divalent amphiphilic
molecules  gain
more entropy, and thus lower the total free energy, by
     staying free. However, as the concentration of divalent surfactant
increases, the gain in electrostatic energy due to
     condensation and  to the hydrophobicity of the
amphiphilic molecules wins over the entropy.

\begin{figure}
\begin{minipage}[t]{70mm}
\includegraphics[width=8cm,height=8cm,angle=-90]{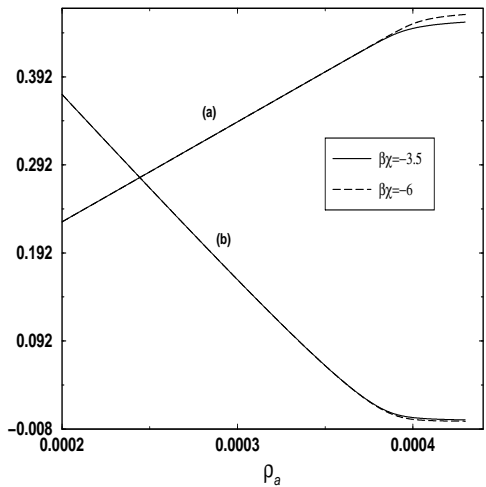}
\caption{Effective binding of amphiphiles
(a)$m_a$ and counterions (b) $m_c$ as
a function of the surfactant concentration $\rho_a$. The
density of DNA and salt are respectively $2\times 10^{-6} M$ and
$10^{-4}M$. The hydrophobicity are $\beta\chi=-3.5$ (solid line) and
$\beta\chi=-6$ (dashed line).}
\label{fig:fig2}
\end{minipage}
\end{figure}
\begin{figure}
\begin{minipage}[t]{70mm}
\includegraphics[width=8cm,height=8cm,angle=-90]{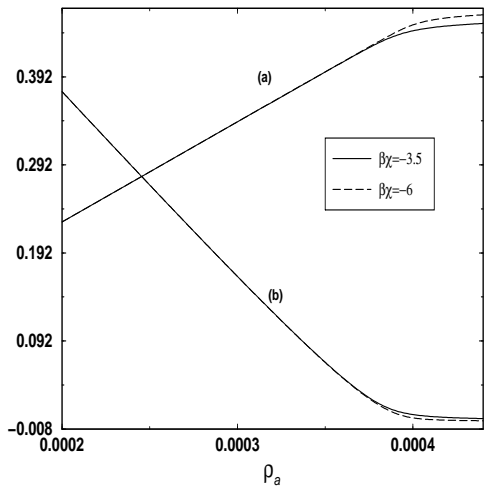}
\caption{The same as Fig.$2$ for $\rho_s=10^{-3}M$.}
\label{fig:fig3}
\end{minipage}
\end{figure}

We note that
     unlike the association with the ionic monovalent
surfactants, which exhibits a large
degree of cooperativity characterized by the
     sharp rise in the surfactant binding fraction, the replacement
of the condensed monovalent counterions by the divalent
     surfactant  proceeds more  smoothly. This result
could have been anticipated a priori.  After the first
amphiphile is associated, the condensation
     of additional molecules is energetically
favored since the buildup of the hydrocarbon density in
the vicinity of a polyion
     helps to exclude water and, thus, reduces the unfavorable
hydrophobic energy of the alkyl tails. This effect competes with
the electrostatic repulsion between the like-charged counterions that
tends to inhibit
it any further association. The repulsion is stronger for
divalent ions than for monovalents and this explain why
the cooperative effect is stronger in the later.
In the case of the divalent amphiphilic molecules the
surfoplex is formed for a density of surfactant
 around  $0.0005M$ as illustrated in Fig. $4$ what  is  much
lower than the one observed in the monovalent case
\cite{Ku98}-\cite{Ku99a}. For comparison with the case
of monovalent surfactant illustrated in refs. \cite{Ku98} for
producing Fig. $4$ we employ the
hydrophobicity factor $\beta \chi =-4$.
\begin{figure}
\begin{minipage}[t]{70mm}

\includegraphics[width=8cm,height=8cm,angle=-90]{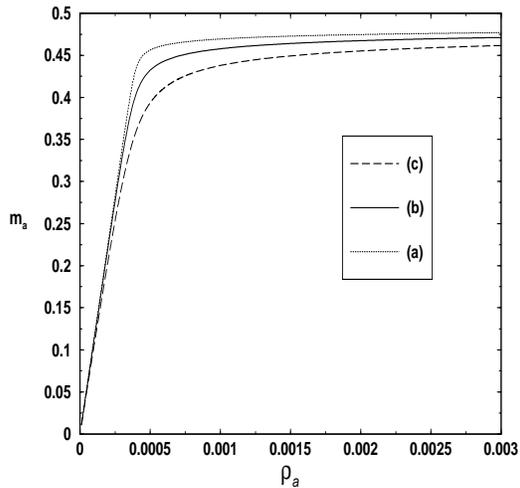}
\caption{Effective binding of divalent surfactant $m_a=n_a/Z$ and
counterions  $m_c=n_c/Z$ as a function of the surfactant
concentration $\rho_a$ for varius salt concentrations: (a) $5mM$,
(b) $18mM$ and (c) $40mM$. The density of DNA is  $2\times 10^{-6}
M$ and  the hydrophobicity is $\beta\chi=-4$.} \label{fig:fig4}

\end{minipage}
\end{figure}

Fig. $5$ and Fig $6$ illustrates the effective charge of the
complex. Due to the simplicity of our model that does not
allow for the association of more than one surfactant molecule
to each charged group along the DNA, there is no charge inversion.
However, the decrease in the charge produced by the divalent
surfactant is larger than the one observed when the
amphiphilic molecules are monovalent for a similar model \cite{Ku98}-\cite{Ku99a}.

\begin{figure}
\begin{minipage}[t]{70mm}
\includegraphics[width=8cm,height=8cm,angle=-90]{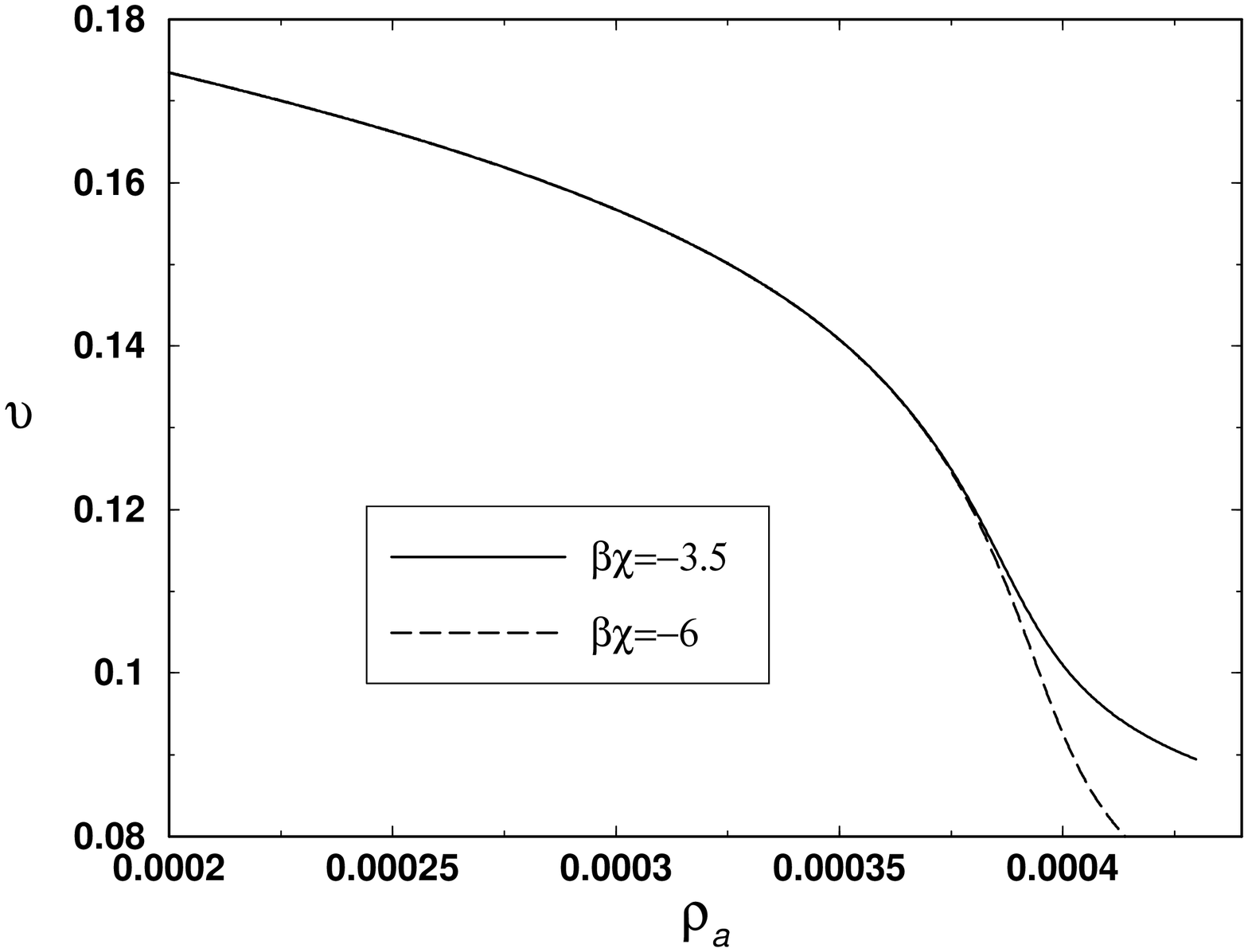}
\caption{The effective charge per charged group of the DNA, $\upsilon=1-2m_a-m_c$
a function of the surfactant concentration $\rho_a$. The
density of DNA and salt are respectively $2\times 10^{-6} M$ and
$10^{-4}M$. The hydrophobicity are $\beta\chi=-3.5$ (solid line) and
$\beta\chi=-6$ (dashed line)
}
\label{fig:fig5}
\end{minipage}
\end{figure}
\begin{figure}
\begin{minipage}[t]{70mm}
\includegraphics[width=8cm,height=8cm,angle=-90]{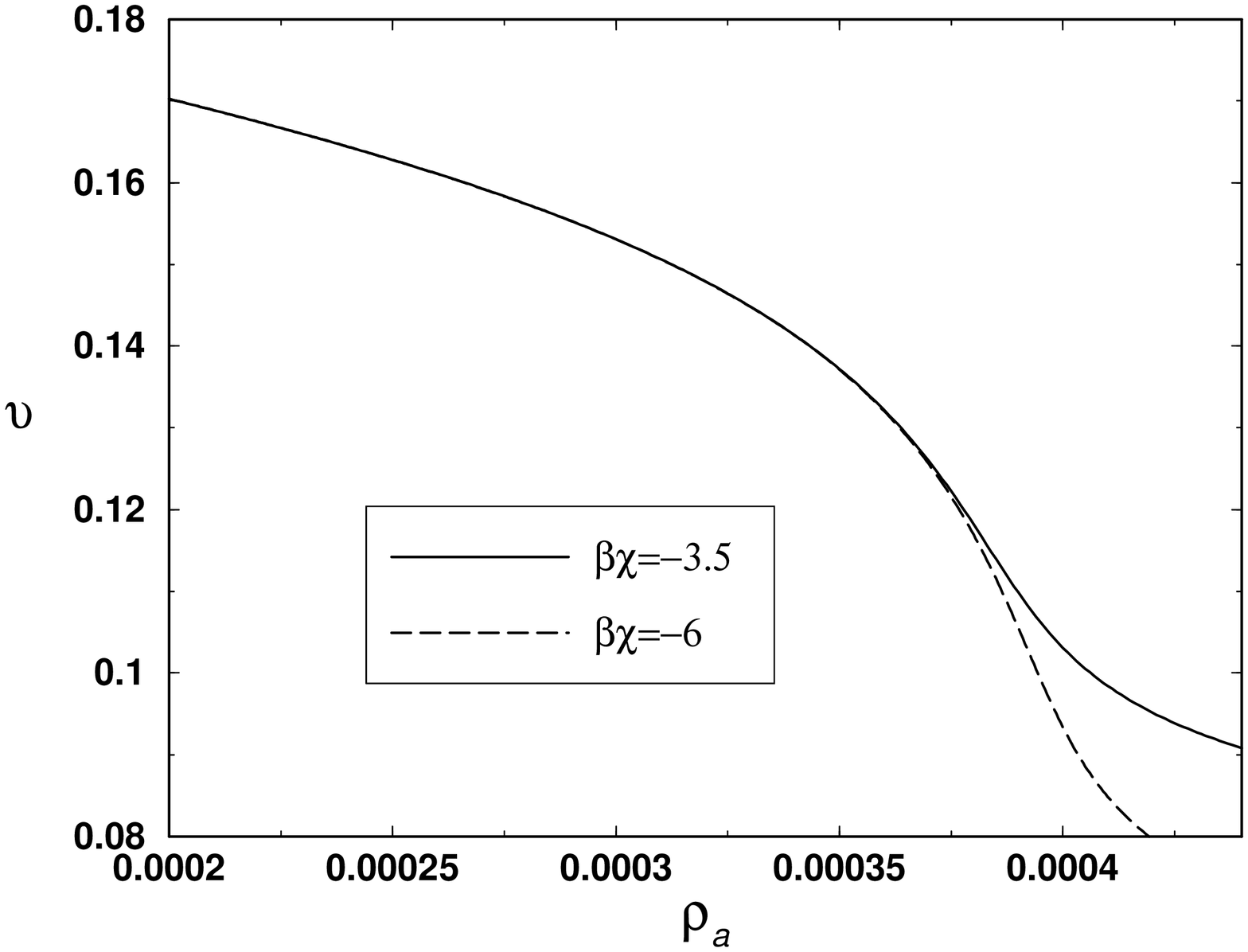}
\caption{The same as Fig. $5$ for $\rho_s=10^{-3}M$}
\label{fig:fig6}
\end{minipage}
\end{figure}

In resume, we have presented a simple theory of DNA,   for monovalent  salt and divalent
surfactant solutions. Our results should be of direct interest to
     researchers working on the design of improved gene delivery systems. In particular, we find that addition of cationic divalent
     surfactants leads to a  cooperative binding. This binding
 happens far below the critical micell
     concentration and far below the concentration for this transition
to happen in the presence of monovalent surfactant.  Until now
   experimental attempts in employing surfactants as  nonviral
agent for transfection have been limited to the use of
monovalent surfactants. The surfoplex produced with monovalent
amphiphilic exhibit poor eficiency. Close to the membrane, most of
the
surfactants disassociate from the DNA and the transfection does
not occurs \cite{Pi89}\cite{Ro91}. We propose the use
of divalent cationic surfactants in the formation
of the complex. Besides being   more strongly connected
to the charged groups of the DNA what will give
more stability to the complex during the transfection, the amount of
surfactant required is lower. Since the surfactant
are toxic to the organism, this should reduce the risk of unnecessary medical complications.

\vspace*{1.25cm}

\noindent{\Large\bf Acknowledgments}

\vspace*{0.5cm} This work was supported by the
brazilian science agencies CNPq, FINEP, PRONEX and CTPETRO

 \end{document}